\documentclass[5p,preprint,sort&compress]{elsarticle}
\usepackage[utf8]{inputenc}
\usepackage{xcolor}
\usepackage{fancyhdr}
\usepackage{amsmath,amssymb}
\usepackage{widetext}
\usepackage{multirow,booktabs,graphicx}
\usepackage{mathtools,feynmp-auto}
\usepackage{multirow}
\usepackage{tikz}
\usepackage{pgfplots}

\graphicspath{{figures/}}

\newcommand{\cth}{c_\theta}
\newcommand{\sth}{s_\theta}

\newcommand\one{\leavevmode\hbox{\small1\normalsize\kern-.33em1}}

\newcommand{\lag}{\mathcal{L}}

\newcommand{\ope}{\mathcal{O}}

\newcommand{\gev}{\text{GeV}}
\newcommand{\tev}{\text{TeV}}

\newcommand{\iab}{\text{ab}^{-1}}
\newcommand{\ifb}{\text{fb}^{-1}}

\def\slashchar#1{\setbox0=\hbox{$#1$}           
   \dimen0=\wd0                                 
   \setbox1=\hbox{/} \dimen1=\wd1               
   \ifdim\dimen0>\dimen1                        
      \rlap{\hbox to \dimen0{\hfil/\hfil}}      
      #1                                        
   \else                                        
      \rlap{\hbox to \dimen1{\hfil$#1$\hfil}}   
      /                                         
   \fi}

\renewcommand{\Vec}{%
  \mathpalette {\overarrow@\vectfill@}}
\def\vectfill@{\arrowfill@\relbar\relbar{\raisebox{-3.81pt}[\p@][\p@]{$\mathord\mathchar"017E$}}}

\newcommand{\be}{\begin{eqnarray*}}
\newcommand{\ee}{\end{eqnarray*}}

\newcommand{\bee}{\begin{eqnarray}}
\newcommand{\eee}{\end{eqnarray}}
\newcommand{\beeq}{\begin{equation}}
\newcommand{\eeeq}{\end{equation}}

\date{}

\title{\vspace*{-2cm}{\scriptsize  \mbox{}\hfill IPPP/20/7, MCNET-20-10}\\[0.5cm]}
\title{Constraining SMEFT operators with associated $h\gamma$ production in\\ Weak Boson Fusion}
\author[1,2]{Anke Biek\"otter}
\author[1,2]{Raquel Gomez-Ambrosio}
\author[1,2]{Parisa Gregg}
\author[1,2]{Frank Krauss}
\author[1]{Marek Sch\"onherr}
\address[1]{Institute for Particle Physics Phenomenology, Durham University, United Kingdom}
\address[2]{Institute for Data Science, Durham University, United Kingdom}

\begin{document}
\begin{fmffile}{feyn-diags}

\begin{abstract}
We consider the associated production of a Higgs boson and a photon in weak boson fusion in the Standard Model (SM) and the Standard Model Effective Theory (SMEFT), with the Higgs boson decaying to a pair of bottom quarks.  Analysing events in a cut-based analysis and with multivariate techniques we determine the sensitivity of this process to the bottom-Yukawa coupling in the SM and to possible CP-violation mediated by dimension-6 operators in the SMEFT. 
\end{abstract}
\maketitle


\section{Introduction}

The observation of the Higgs boson in 2012~\cite{Aad:2012tfa,Chatrchyan:2012xdj} initiated intense efforts to measure its properties in a wide range of production and decay processes, to either confirm it as the Higgs boson predicted by the Standard Model, or to catch first glimpses of new physics beyond it. To date, no significant deviation has been found~\cite{Aad:2019mbh,Sirunyan:2018koj} and the Standard Model appears to be a robust and healthy theory. 
As a consequence the focus has shifted from the discovery of signals of new physics models to model-independent constraints on experimentally allowed deviations from Standard Model predictions.  

In this work, we study the associated production of a Higgs boson with a photon in weak boson fusion (WBF), manifesting itself in a final state consisting of the two bosons and two forward jets.  This process was first proposed as a possibly interesting Higgs boson production channel in~\cite{ABBJJ_Maltoni, Asner:2010ve}~\footnote{In this process, the $hWW$ vertex is even more important than the $hZZ$ coupling, compared to WBF $h$ production.} .  With the Higgs boson decaying into two $b$ quarks, the additional photon efficiently suppresses otherwise dominant QCD backgrounds. 
The ATLAS collaboration has studied this channel in~\cite{Aaboud:2018gay} with a boosted decision tree at $30.6\,\ifb$ and found a signal significance of $1.4\sigma$.  
Using a cut-based analysis and contrasting it with multivariate techniques we analyse the potential of this channel for an independent measurement of the bottom-Yukawa coupling at higher luminosities.  

We further investigate the impact of possible effects of beyond the Standard Model physics in WBF $h\gamma$ production, using the language of effective dimension-six operators from the Standard Model Effective Theory (SMEFT) ~\cite{Buchmuller:1985jz, Georgi:1994qn, Grzadkowski:2010es, Alonso:2013hga,Brivio:2017vri}.  
Wilson coefficients of SMEFT operators relevant in Higgs physics have been constrained through various channels including WBF, for example in~\cite{Masso:2012eq,Falkowski:2015jaa,Brooijmans:2014eja,Trott:2014dma,Falkowski:2014tna,Belanger:2013xza,Giardino:2013bma,Banerjee:2013apa,Dumont:2013wma,Bechtle:2014ewa,Cheung:2014noa,Ellis:2014dva,Englert:2015hrx,Flament:2015wra,Bian:2015zha,Buchalla:2015qju,Fichet:2015xla,Corbett:2015ksa,Reina:2015yuh,Butter:2016cvz,deBlas:2016ojx,deBlas:2016nqo, deBlas:2017wmn,Ellis:2018gqa,Almeida:2018cld,Biekotter:2018rhp}.
Here, we advocate to also use WBF production of the $h\gamma$ final state as an additional, independent constraint.  While the kinematic structure of the interactions induced by $CP$-even operators renders WBF Higgs boson production the by far preferred process, we focus in particular on $CP$-odd operators in the gauge-Higgs sector of SMEFT.  They exhibit comparable sensitivity in both WBF $h$ and $h\gamma$ production.
In addition, the limits on this set of operators provides important constraints an additional sources of $CP$ violation, necessary to describe, for example, electroweak baryogenesis~\cite{Sakharov:1967dj,Kuzmin:1985mm,Shaposhnikov:1987tw,Nelson:1991ab,Morrissey:2012db}. The $CP$-odd dimension-6 EFT operators considered in our analysis have been studied and constrained in Higgs boson~\cite{Ferreira:2016jea,Brehmer:2017lrt,Bernlochner:2018opw,Englert:2019xhk,Cirigliano:2019vfc} and diboson production processes~\cite{Kumar:2008ng,Dawson:2013owa,Gavela:2014vra,Azatov:2019xxn}. Our study further extends this list of relevant signatures and proposes a sensitive experimentally accessible observable, which we use to constrain two of the $CP$-odd operators of the dimension-6 EFT basis.

\section{Signal and backgrounds in the Standard Model and determination of the $b$-Yukawa coupling}
\label{sec:simulation}

\subsection{Process simulation}
For our study we assume $\sqrt{s} = 13$ TeV throughout. 
The signal process ($h\gamma$ production in association with two jets at $\ope(\alpha^4)$, both in the SM and in SMEFT) is simulated with \texttt{MadGraph5, v2.6.6}~\cite{MadGraph} at leading order (LO) and with the default \texttt{NNPDF23\_NLO} parton distribution function~\cite{Ball:2013hta}. 
\texttt{PYTHIA 8.2}~\cite{Sjostrand:2014zea} models secondary emissions through parton showering, performs hadronization and adds the underlying event; it also decays the Higgs boson into the $b$-quarks.
We select the WBF topology through the usual invariant mass cut on the tagging jets~$m_{jj}$; all jets, at both parton and hadron level, are defined through the anti-kT algorithm~\cite{Cacciari:2008gp} with $R=0.4$. In the following, the indices $j$ and $b$ refer to the light and $b$-jets.
At generation level the following parton-level cuts are applied to final-state transverse momenta~$p_T$ and pseudo-rapidities~$\eta$ 
\begin{equation}\label{eq:SMgencuts}
  \begin{split}
    &\;p_{T,j} > 30\, \gev , \quad |\eta_j| < 5., \\
    &\;p_{T,\gamma} > 20\, \gev , \quad |\eta_\gamma| < 2.5, \\
    &\;\Delta R_{\gamma j}>0.4 ,   \quad m_{jj} > 1200\,\gev.
  \end{split}
\end{equation}
The combination of these cuts ensures that non-WBF contributions
(gluon fusion, $tth$ and $Vh$) to the signal are negligible at the 10\% level~\cite{Aaboud:2018gay}.\\

All irreducible background processes are simulated at LO using \texttt{Sherpa-2.2.7}~\cite{Gleisberg:2008ta} with the default \texttt{NNPDF30\_NNLO} parton distribution function~\cite{Ball:2014uwa} from \texttt{LHAPDF 6.2.1}~\cite{Buckley:2014ana}; matrix elements are calculated with \texttt{COMIX}~\cite{Gleisberg:2008fv} and jets are parton showered with \texttt{CSSHOWER++}~\cite{Schumann:2007mg}~\cite{Nagy:2005aa}. For hadronisation etc.\ we use the Sherpa default settings. 

Background contributions to the signal final-state feature the direct production of $b$-jets in the simulation, necessitating additional generation-level cuts.  We consider the following processes:
\begin{itemize}
    \item Continuum production of a $b$-jet pair, two light jets and a photon, $b\bar bjj\gamma$. In particular,
    we consider $\ope(\alpha_s^4\alpha)$ contibutions which we denote QCD and electroweak (EW) $Z\gamma j j$ production with the Z boson decaying to $b$-quarks, with the following 
    additional cuts on the $b$'s:
    \begin{equation}
      \begin{split}
        &\;p_{T,b} > 20\,\gev, \quad  m_{bb} \in [90,  \, 200]\,\gev, \\ 
        &\;\Delta R_{\gamma b}>0.4,\quad \Delta R_{jb}>0.4\,,
      \end{split}
    \end{equation}
    We have explicitly checked that the contributions from $\ope(\alpha_s^2 \alpha^3)$ are negligible at the $5\%$ level and $\ope(\alpha_s^3 \alpha^2)$ as well as $\ope(\alpha_s \alpha^4)$ contribute less than $1\%$ each.
    \item $t\bar{t}\gamma$ production and single top production with an associated photon. 
    For the $t\bar{t}\gamma$ and single top processes we force the decay of the $W^\pm$ boson to light quarks.  We do not apply specific cuts on the decay products of the on-shell top quarks, but we require, again, 
    \begin{align}
    \Delta R_{\gamma j}>0.4
    \end{align}
    for the single-top processes.
\end{itemize}

\subsection{Extracting the signal}
In the initial analysis with \texttt{Rivet 2.7.0}~\cite{Buckley:2010ar} we apply the following baseline cuts to all signal and background processes:
\begin{enumerate}
    \item We require an isolated photon with 
    \begin{align}
        p_{T\gamma}>20\, \gev, \quad|\eta_\gamma|<2.5 
        \label{eq:cutflow1}
    \end{align}
    and the isolation given by
    \begin{align}
    \sum\limits_{i, \Delta R_{i\gamma}<0.4} p_\perp^i < 10\,\gev.
    \end{align}
    \item We require exactly two light jets and two $b$-jets, 
    \begin{align}
        N_\text{jets} = N_\text{b-jets} = 2,
    \end{align}
    where both are defined with the anti-kT algorithm with R=0.4 and
    \begin{equation}
      \begin{split}
        &\;p_{Tj} > 40\, \gev, \quad p_{Tb} > 30 \,\gev,\\
        &\:|\eta_{j_1}|< 4.5, \quad |\eta_b|<2.5 \, .
      \end{split}
    \end{equation}
    We assume perfect b-tagging efficiency. 
    \item To select the WBF topology, we cut on the invariant light jet mass and the pseudo-rapidity 
    difference of the light jets
    \begin{align}
        m_{jj} > 1500 \,\gev, \quad \Delta \eta_{jj} > 4.5 \, .
    \end{align}
    \item We require the invariant $b$-jet mass to be close to the Higgs mass
    \begin{align}
        m_{bb} \in [100,\, 140]\, \gev \, .
        \label{eq:baseline_cuts}
    \end{align}
    This finalizes our baseline selection which we will use in the multivariate analysis in Section~\ref{subsec:Yukawa}.
    \item To allow for a fair comparison between a cut-and-count approach and the multivariate analysis below, 
    we apply the following additional cuts in our cut-and-count analysis
    \begin{equation}\label{eq:final_cuts}
      \begin{split}
        &\;|\eta_{j_1}|>1.5\,, \quad |\eta_{j_2}|>2\,,\\
        &\;\eta^\text{cen}_{\gamma bb}, \, \eta^\text{cen}_{\gamma}\,, \, \eta^\text{cen}_{bb} < 0.5\,, \\
        &\;\quad m_{jj}>2000\,\gev,
      \end{split}
    \end{equation}
    where the centralities $\eta^\text{cen}_x$ relative to the WBF tagging jets are defined as 
    \begin{align}
      \eta^\text{cen}_x =\, \left| \frac{\eta_x - \frac{\eta_{j_1}+\eta_{j_2}}{2}}{\eta_{j_1}-\eta_{j_2}} \right| . 
    \end{align}
\end{enumerate}

\begin{figure}[t!]
\centering
	\includegraphics[width=.45\textwidth]{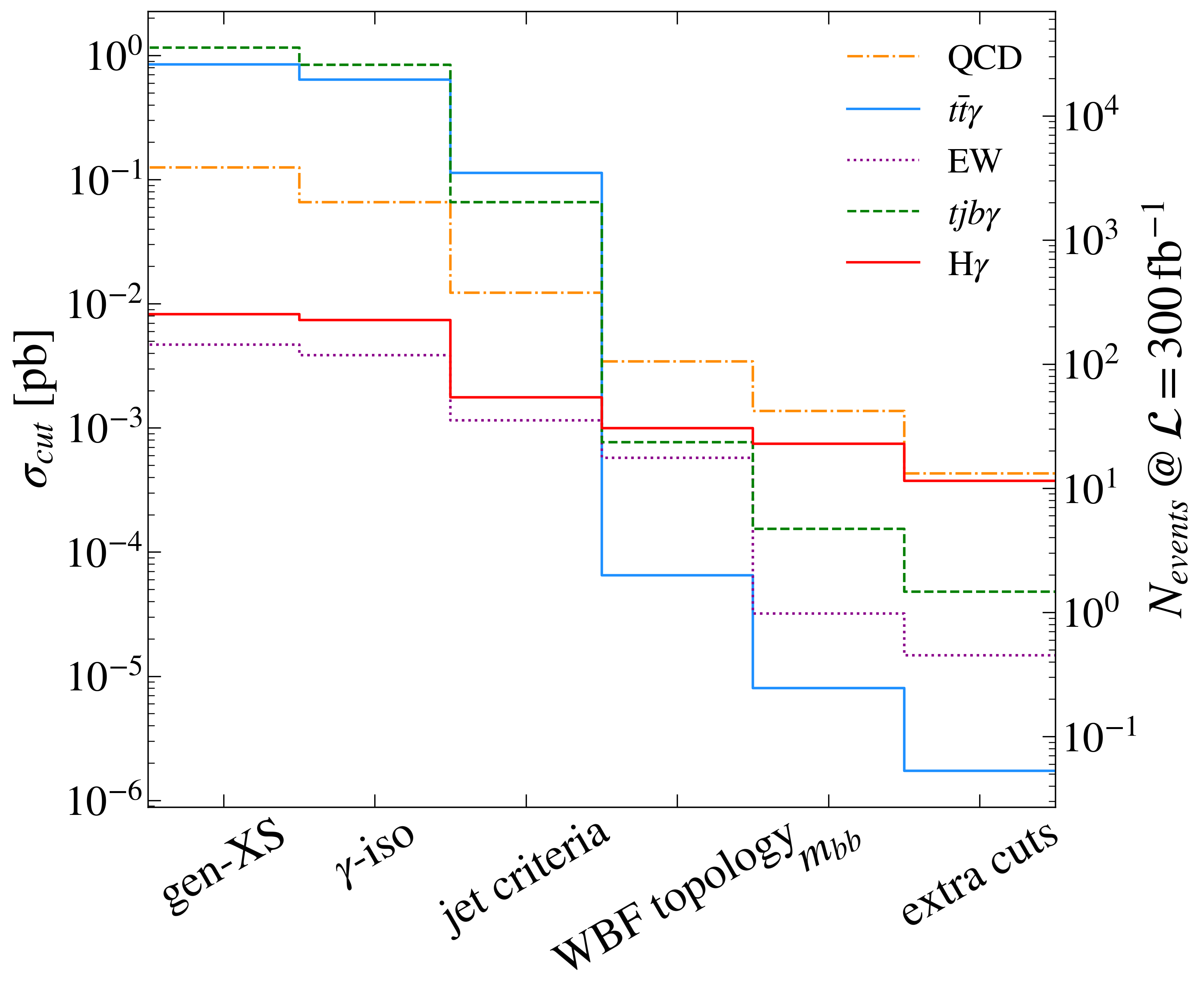}
	\caption{Cross section after different cuts in our cutflow, as given in
	Eqs.~\eqref{eq:cutflow1}-\eqref{eq:final_cuts}. On the right axis, we display the number of 
	events for an integrated luminosity of $300\,\ifb$. Be aware that we applied 
	stronger cuts on the QCD and EW backgrounds at generator level which explains 
	their lower generator-level cross section compared to the top backgrounds. }
	\label{fig:cutflow}
\end{figure}
The signal and background process cutflow is shown in Fig.~\ref{fig:cutflow}.  The baseline set of cuts, Eq.~\eqref{eq:baseline_cuts}, reduces the contribution from $t\bar{t}\gamma$ and single top processes by six and four orders of magnitude, respectively, whilst only loosing one order of magnitude in the signal. With the top-based backgrounds irrelevant after cuts, the dominant background contribution for associated $h\gamma$ production stems from the continuum QCD process.

After the final cuts in Eq.~\eqref{eq:final_cuts}, we reach a signal-over-background ratio of $S/B = 0.8$ 
in our cut-and-count analysis.
We translate this into a CL$_s$ limit~\cite{Read:2002hq} on the signal strength
\begin{align}
  \mu=\frac{\sigma (\text{pp} \rightarrow h j j \gamma) \text{ BR}( h \rightarrow b \bar{b})}
{ \sigma^\text{SM} (\text{pp} \rightarrow h j j \gamma) \text{ BR}^\text{SM}( h \rightarrow b \bar{b})}  
\end{align}
using the CL$_s$ limit setting implementation in \texttt{CheckMATE} \cite{Dercks:2016npn}.
The resulting limits are $\mu < 1.1$ for $\lag_\text{int}=30.6\,\ifb$ at $95\%$ CL ($\mu < 0.4$ for $\lag_\text{int}=300\,\ifb$ and $\mu < 0.3$ for $\lag_\text{int}=3000\,\ifb$) assuming negligible systematic uncertainties.

\subsection{Determination of the $b$-Yukawa coupling}
\label{subsec:Yukawa}
Since the coupling of the photon to quarks and gauge bosons as well as 
gauge-boson--quark couplings are very precisely known, 
the WBF~$h\gamma$ signature will allow us to independently 
constrain the Higgs Yukawa coupling to the $b$-quark in the WBF topology. 

To further increase the sensitivity to our search with respect to the final cuts in Eq.~\eqref{eq:final_cuts}, we perform a 
multivariate analysis with \texttt{TMVA}~\cite{tmva} in \texttt{Root~6.22}~\cite{Brun:1997pa}. 
We find the optimal signal regions -- dependent on the luminosity -- by passing the events satisfying the baseline selection cuts 
of Eq.~\eqref{eq:baseline_cuts} to a Boosted Decision Tree (BDT).
Our pre-selection cuts are much stronger than the ones included in the experimental analysis in 
Ref.~\cite{Aaboud:2018gay}. In particular, our cut on the invariant mass of the tagging jets $m_{jj}> 1500 \,\gev$ 
is much tighter than the ATLAS constraint of $m_{jj}> 800 \,\gev$, 
thereby effectively negating any effect of the $Z\gamma$ (EW) contribution.
We generate $N=200$~trees with a maximum depth of~$3$ 
and set the minimum node size to $6\%$ to avoid over-training. 
As our input variables, we choose the $p_T$ and $\eta$ of all final-state particles, as well 
as 
\begin{equation}
  \begin{split}
    &\;
    m_{jj}, \quad 
    \Delta\eta_{jj}, \quad 
    \Delta\phi_{jj}, \quad 
    \Delta R_{\gamma, j_1}, \quad 
    \Delta R_{\gamma, j_2}, \quad
    \\
    &\;
    m_{bb}, \quad 
     \Delta\eta_{bb},  \quad 
    \Delta\phi_{bb}, \quad 
    \Delta R_{\gamma, b_1}, \quad 
    \Delta R_{\gamma, b_2}, \quad 
    \\
    &\;
    p_{T,bb}, \quad 
    \eta_{bb}, \quad
    \\
    &\;
    m_{bb\gamma} , \quad
    \Delta \eta_{\gamma,bb} , \quad 
    \Delta\phi_{\gamma, bb}, \quad
    \Delta R_{\gamma, bb},  \quad 
    \\
    &\;
    \eta^\text{cen}_{\gamma bb}, \quad 
    \eta^\text{cen}_{\gamma }, \hspace*{1.6ex}
    \eta^\text{cen}_{bb} \, .
  \end{split}
\end{equation}
As expected, the variable that is most often used by the BDT is $m_{bb}$ which is peaked around the Higgs mass for the signal, but flat for the dominant QCD  background. We have checked explicitly that after the cuts on the BDT classifier~$\chi_\text{BDT}$ used  for our limit setting we do not focus on a range of $m_{bb}$ below the experimental detector resolution,
cf.\ Fig.~\ref{fig:mbb}. 
All other input observables are less important individually, but collectively contribute much more than $m_{bb}$. 
Removing $m_{bb}$ as an input variable altogether reduces the efficiency of the signal classification at a fixed 
background efficiency of $10\%$ by about $10\%$. 
In Fig.~\ref{fig:ROC} we contrast the BDT ROC curve with the cut-\-and-\-count analysis efficiency. The BDT analysis clearly outperforms the cut-\-and-\-count approach for this rather complicated final state. 
\begin{figure}[t!]
	\hspace*{4pt}
	\includegraphics[height=.4\textwidth]{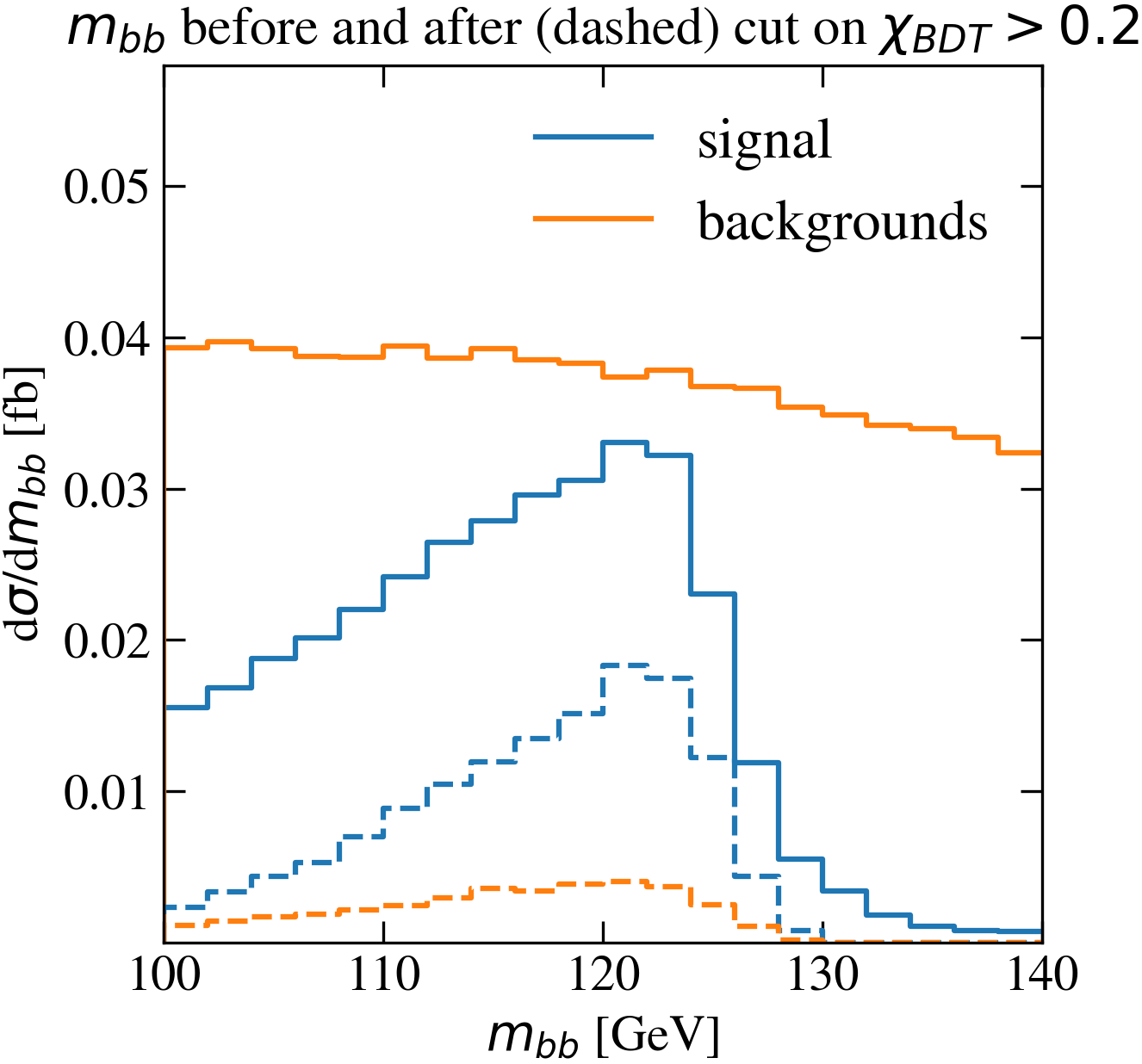}
	\caption{Distribution of the invariant mass of the $b$-jet pair~$m_{bb}$ before (solid lines)
	and after (dashed lines) a cut on the BDT classifier of $\chi_\text{BDT}>0.2$.
        }
	\label{fig:mbb}
\end{figure}

\begin{figure}[t!]
	\hspace*{8pt}
	\includegraphics[height=.4\textwidth]{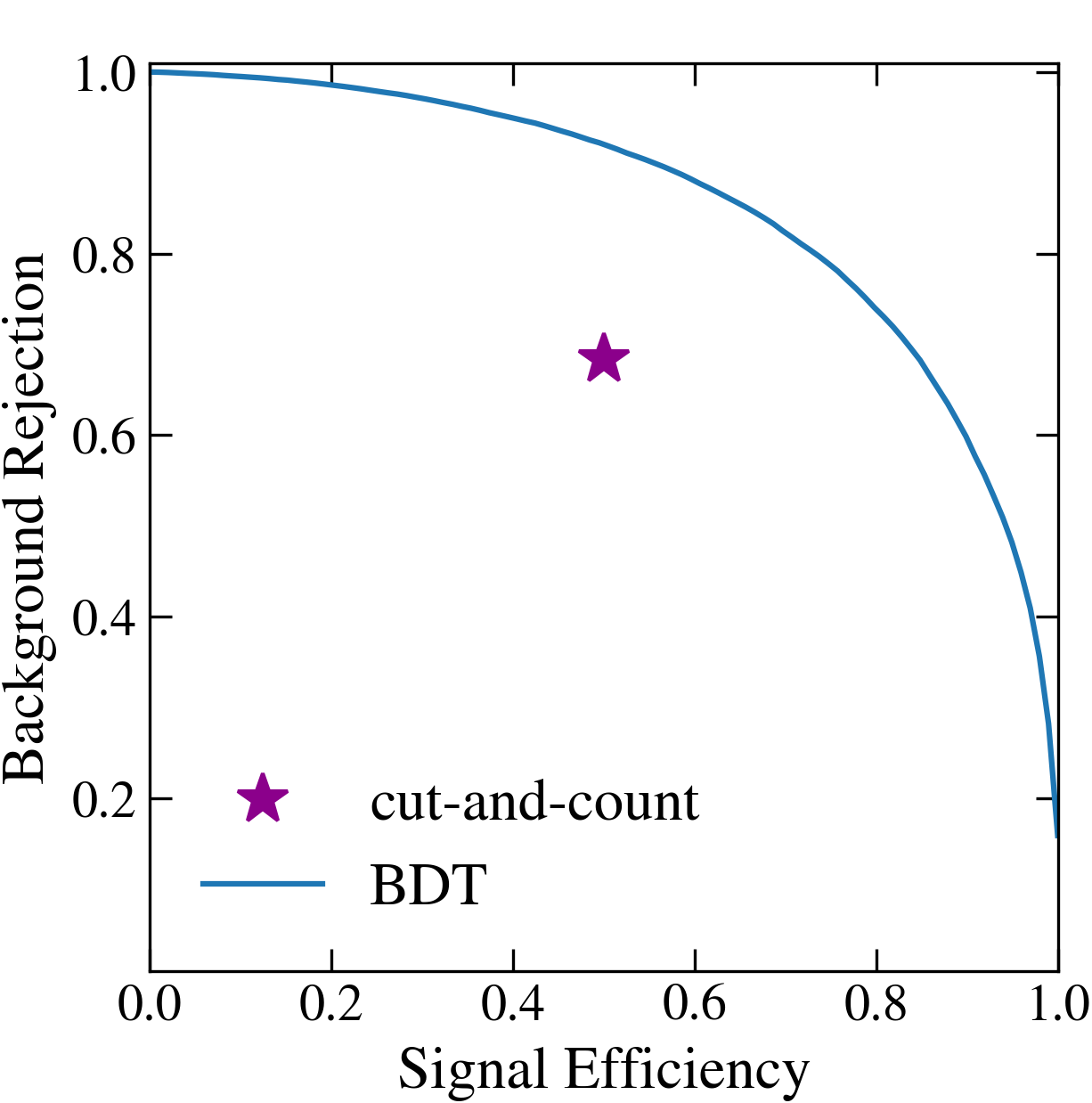}
	\caption{Receiver operating characteristic curve for the BDT analysis. The asterisk marks the 
	signal and background efficiencies after the cuts in Eq.~\eqref{eq:final_cuts} compared 
	to the baseline cuts, Eq.~\eqref{eq:baseline_cuts}.}
	\label{fig:ROC}
\end{figure}

For a given luminosity, we choose the BDT classifier cut which minimizes 
the CL$_s$ limit on the WBF~$h\gamma$ signal strength~$\mu$.
In our limit setting, we assume statistical uncertainties to be dominant and therefore 
neglect systematic uncertainties. 
For $\lag_\text{int}=30.6\,\ifb$, the resulting $95\%$~CL$_s$ limit is $\mu < 0.8$ for a cut 
on the BDT classifier of $\chi_\text{BDT}>0.1$. After this cut we are left with $6.2$ expected signal and $2.8$ expected background events. 
At $\lag_\text{int}=300\,\ifb$ and $\lag_\text{int}=3000\,\ifb$ these limits will increase to $\mu < 0.25$ (for $\chi_\text{BDT}>0.2$) 
and $\mu < 0.1$ (for $\chi_\text{BDT}>0.2$), 
respectively. This clearly indicates that an observation of the decay channel $h \rightarrow b \bar{b}$
will be possible at the HL-LHC.
Notice again, however, that the calculation assumes negligible systematic uncertainties
which will no longer be true at higher luminosities. Assuming a $50\%$ systematic uncertainty on the 
backgrounds, the above limits weaken to $\mu < 0.9, \, 0.3, \, 0.15$ at the $95\%$~CL for integrated luminosities 
of $\lag_\text{int}=30.6\,\ifb, \, 300\,\ifb, \, 3000\,\ifb$ respectively.

\section{EFT analysis}  
 \label{sec:EFT}
\subsection{Selection of operators}
We continue with an analysis of potential BSM effects affecting the signal.  Effects are, as usual, parametrized in terms of an effective Lagrangian, truncated at dimension-six~\cite{Weinberg:1978kz,Georgi:1985kw,Donoghue:1992dd,Buchmuller:1985jz, Georgi:1994qn, Grzadkowski:2010es, Alonso:2013hga,Brivio:2017vri}, 
\begin{align}
    \lag_\text{SMEFT} = \lag_\text{SM} + \sum_i \frac{c_i \, \ope_i^{(6)}}{\Lambda^2} \, ,
\end{align}
where the $c_i$ are the Wilson coefficients. They correspond to the operators~$\ope_i$ in the Warsaw basis~\cite{Grzadkowski:2010es} which are suppressed by inverse powers of the new physics scale~$\Lambda$.  
Due to the relatively small cross section of our signal, 
many if not all of the Wilson coefficients of these operators will be constrained by other processes 
before our signal starts to become sensitive. 
In addition, in some other processes, tri-linear boson couplings (such as $VVV$ or $VVh$) experience high-momentum enhancement which is not the case for four-boson interactions, like $WW\gamma h$.  However, our signal can provide an independent probe of paradigms underlying the construction of the effective field theory framework and may also help in lifting possible degeneracies in global fits.

\begin{figure}[t!]
    \centering
      \begin{fmfgraph*}(90,40)  
        \fmfset{arrow_len}{2.5mm}
        \fmfleft{i1,i2}  
        \fmfright{f1,fh,fb,f2}  
        \fmf{fermion}{v1,i1}    
        \fmf{fermion}{i2,v2}   
        \fmf{boson}{v1,v3,v4,v2} 
        \fmf{fermion}{f1,v1}    
        \fmf{fermion}{v2,f2}  
        \fmffreeze 
        \fmf{dashes}{v3,fh} 
        \fmf{boson}{v4,fb} 
      \end{fmfgraph*}
      \quad
      \begin{fmfgraph*}(90,40)  
        \fmfset{arrow_len}{2.5mm}
        \fmfleft{i1,i2}  
        \fmfright{f1,fdummy,fh,fb,f2}  
        \fmf{fermion}{f1,v5}    
        \fmf{fermion}{v4,f2}
        \fmf{fermion,tension=1.5}{v5,v1}    
        \fmf{fermion,tension=1.5}{v2,v4} 
        \fmf{fermion,tension=0.8}{v1,i1}    
        \fmf{fermion,tension=0.8}{i2,v2}  
        \fmf{boson}{v1,v3,v2}
        \fmffreeze 
        \fmf{dashes}{v3,fh} 
        \fmffreeze
        \fmf{phantom}{i2,v2,v4,f2} 
        \fmf{phantom}{v5,fdummy} 
        \fmf{boson}{v4,fb} 
      \end{fmfgraph*}
      \\[5mm]
      \begin{fmfgraph*}(90,40)  
        \fmfset{arrow_len}{2.5mm}
        \fmfleft{i1,i2}  
        \fmfright{f1,fh,fb,f2}  
        \fmf{fermion}{v1,i1}    
        \fmf{fermion}{i2,v2}   
        \fmf{boson}{v1,v3,v2} 
        \fmf{fermion}{f1,v1}    
        \fmf{fermion}{v2,f2}  
        \fmffreeze 
        \fmf{dashes}{v3,fh} 
        \fmf{boson}{v3,fb} 
        \fmfblob{.1w}{v3}
      \end{fmfgraph*}
      \quad
      \begin{fmfgraph*}(90,40)  
        \fmfset{arrow_len}{2.5mm}
        \fmfleft{i1,i2}  
        \fmfright{f1,fh,fb,f2}  
        \fmf{fermion}{v1,i1}    
        \fmf{fermion}{i2,v2}   
        \fmf{boson}{v1,v3,v4,v2} 
        \fmf{fermion}{f1,v1}    
        \fmf{fermion}{v2,f2}  
        \fmffreeze 
        \fmf{dashes}{v3,fh} 
        \fmf{boson}{v4,fb} 
        \fmfblob{.1w}{v3}
      \end{fmfgraph*}
      \\[5mm]
      \begin{fmfgraph*}(90,40)  
        \fmfset{arrow_len}{2.5mm}
        \fmfleft{i1,i2}  
        \fmfright{f1,fdummy,fh,fb,f2}  
        \fmf{fermion}{f1,v5}    
        \fmf{fermion}{v4,f2}
        \fmf{fermion,tension=1.5}{v5,v1}    
        \fmf{fermion,tension=1.5}{v2,v4} 
        \fmf{fermion,tension=0.8}{v1,i1}    
        \fmf{fermion,tension=0.8}{i2,v2}  
        \fmf{boson}{v1,v3,v2}
        \fmffreeze 
        \fmf{dashes}{v3,fh} 
        \fmffreeze
        \fmf{phantom}{i2,v2,v4,f2} 
        \fmf{phantom}{v5,fdummy} 
        \fmf{boson}{v4,fb} 
        \fmfblob{.1w}{v3}
      \end{fmfgraph*}
    \caption{Example diagrams for WBF Higgs production in association with a photon in the SM (top row) and in the EFT (centre and bottom row).}
    \label{fig:feyn_WBF}
\end{figure}
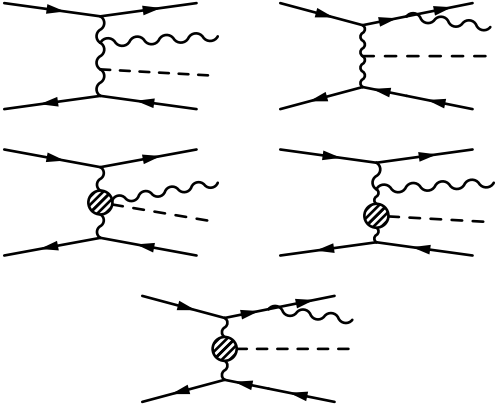

There are many operators contributing to our signal process, for example through their modifications of fermion-gauge or Higgs-gauge couplings.  They can be tested (and better constrained) in WBF without an additional photon or other processes. 
Here, we will focus on operators which lead to contact interactions of three gauge bosons and a Higgs boson as depicted in the centre left diagram in Fig.~\ref{fig:feyn_WBF}, and the gauge-related subsets of effective three-point interactions. The diagrams with four-point interactions have the advantage of being suppressed by only two t-channel $W$~propagators, compared to the SM which is suppressed by three $t$-channel $W$~propagators when the photon is radiated from the $W$ bosons and not from one of the quark lines. We can enhance their relative importance by requiring a large $\Delta R_{\gamma j}$ separation between the photon and the jets.  A contact interaction of three gauge bosons and a Higgs boson exists for the following operators 
\begin{align}
    \ope_{HW} &= H^\dagger H W^I_{\mu \nu} W^{I \mu \nu} &\; 
    \ope_{H\tilde{W}} &= H^\dagger H \tilde{W}^I_{\mu \nu} W^{I \mu \nu} \nonumber \\
    \ope_{HWB} &= H^\dagger \tau^I H W^I_{\mu \nu} B^{\mu \nu} &\; 
    \ope_{H\tilde{W}B} &= H^\dagger  \tau^I H \tilde{W}^I_{\mu \nu} B^{\mu \nu} \, .
\end{align}
The four-point interaction of three gauge bosons and a Higgs boson $WW\gamma h$ which results from these operators structurally looks like\\
\begin{widetext}
\hspace*{0.05\textwidth}
\begin{minipage}{.2\textwidth}
\begin{fmfgraph*}(50,50)  
        \fmfset{arrow_len}{2.5mm}
        \fmfleft{i1,i2}  
        \fmfright{o1,o2}  
        \fmf{boson,label=$p_1$,label.side=left}{v,i2}
        \fmf{boson,label=$p_2$,label.side=left}{i1,v}
        \fmf{boson,label=$p_\gamma$,label.side=left}{v,o1} 
        \fmf{dashes}{v,o2} 
        \fmfblob{.25w}{v}
        \fmflabel{$W^-_\nu$}{i1}
        \fmflabel{$W^+_\mu$}{i2}
        \fmflabel{$\gamma_\rho$}{o1}
        \fmflabel{$h$}{o2}
      \end{fmfgraph*}
      \quad
\end{minipage}
\begin{minipage}{.7\textwidth}
\begin{equation}\label{eq:structures}
  \begin{split}
    & c_{HWB}\frac{2i e v}{\Lambda^2} \frac{\sth}{\cth} \, \left(p_\gamma^{\mu} g_{\alpha \nu} - p_\gamma^{\nu} g_{\alpha \mu} \right) \\
    & c_{HW}\frac{-4i e v}{\Lambda^2}\, \left( 
      g_{\alpha \mu} (p_\gamma- p_1)^{\nu}  
    + g_{\alpha \nu} (p_2 - p_\gamma)^{\mu}  
    + g_{\mu \nu} (p_1- p_2)^{\alpha}  \right)
    \\
    & c_{HB\tilde{W}}\frac{- 2i e v}{\Lambda^2} \frac{\sth}{\cth} \,
     \epsilon_{\alpha \nu \mu \rho}\, p_\gamma^\rho
     \\
    & c_{H\tilde{W}}\frac{4i e v}{\Lambda^2}\, 
    \epsilon_{\alpha \nu \mu \rho} \,
    \left(  p_\gamma + p_1 + p_2\right)^{\rho} 
  \end{split}
\end{equation}
\end{minipage}
\end{widetext}
The Lorentz structure of the four-point interaction resulting from $\ope_{HW}$ is identical to the one from the SM $WW\gamma$ vertex. For this operator, the EFT and SM diagrams differ only by the additional t-channel $W$~propagator in the SM case.
The three-point $VVh$ counterpart of the above operator has an additional momentum enhancement from derivatives in the $W_{\mu\nu}$ field 
strength tensors. For comparison, we show the structures of the 
$WWh$ interaction resulting from the operator $\ope_{HW}$ and its $CP$-odd counterpart $\ope_{H\tilde{W}}$. The operators $\ope_{HWB}$ and $\ope_{H\tilde{W}B}$ 
contribute to $hZ\gamma$, $h\gamma\gamma$ and $hZZ$ couplings only.
These couplings will be less relevant for our study because they do not allow for the 
photon to be radiated off the t-channel propagators and 
the contribution of diagrams in which the photon is 
radiated off a jet is suppressed by the cuts on the angle between 
the photon and the jets. \\\vspace*{3mm}

\begin{minipage}{.4\columnwidth}
\hspace{6mm}
\begin{fmfgraph*}(50,50)  
        \fmfset{arrow_len}{2.5mm}
        \fmfleft{i1,i2}  
        \fmfright{o}  
        \fmf{boson,label=$p_1$,label.side=left}{v,i2}
        \fmf{boson,label=$p_2$,label.side=left}{i1,v}
        \fmf{dashes}{v,o} 
        \fmfblob{.25w}{v}
        \fmflabel{$W^-_\nu$}{i1}
        \fmflabel{$W^+_\mu$}{i2}
        \fmflabel{$h$}{o}
      \end{fmfgraph*}
\end{minipage}
\begin{minipage}{.55\columnwidth}
\begin{equation}
  \begin{split}
    & c_{HW}\frac{4i e v}{\Lambda^2}\, \left( 
      p_1^\nu p_2^\mu - g_{\mu \nu} \, p_1\cdot p_2 \right)
    \\
    & c_{H\tilde{W}}\frac{4i e v}{\Lambda^2}\, 
    \epsilon_{\mu \nu \rho \delta} \,
     p_1^{\rho}  p_2^{\delta} 
  \end{split}
\end{equation}
\end{minipage}\\\vspace*{3mm}

Events for the EFT signal contributions have been generated with the
SMEFTsim implementation~\cite{SMEFTsim} of the Warsaw basis 
with \texttt{MadGraph}~\cite{MadGraph}, neglecting dimension-six
squared terms. 
We apply the same cuts as for the SM signal. This includes the cuts 
in Eq.~\eqref{eq:SMgencuts} on generator level, as well as 
the baseline selection cuts in Eq.~\eqref{eq:baseline_cuts} after parton 
showering and hadronization. 
After these cuts,
we can parametrize the WBF~$h\gamma$ cross section as 
\begin{equation}\label{eq:cross_section}
  \begin{split}
    \lefteqn{\left.\frac{\sigma^{(\text{LO})}_{\text{SM}+1/\Lambda^2}}{\sigma^{(\text{LO})}_\text{SM}}\right|_\text{cuts}-1}\\
    &\; 
    = 10^{-3}\,\cdot\left(\frac{1\,\tev}{\Lambda}\right)^2\, \cdot
    \left[\vphantom{\int}
    -44 \, c_{HW}
    -240 \, c_{HWB}
    \right]\,.
  \end{split}
\end{equation}
The $CP$-odd operators do not contribute to the total
cross section on the level of interferences of the EFT
with the SM only. We will see in the next section how they can still have 
observable consequences for angular distributions of final 
state particles. 

\subsection{$CP$ structure of the EFT and observable consequences}
The vertex structures of the operators $\ope_{H\tilde{W}}$ and $\ope_{H\tilde{W}B}$, 
as given in Eq.~\eqref{eq:structures}, lead to $CP$ violation.
Currently, the best constraints on these operators in the Higgs sector 
come from observables in WBF and the Higgs decay $h \rightarrow ZZ \rightarrow 4 \ell$ 
respectively~\cite{Bernlochner:2018opw}. 
For our process, we can construct $CP$-sensitive observables 
from combinations of scalar products and cross products of the momenta of the final state particles. 
As we have four particles in the final state, there are multiple ways to 
combine the momenta. Scanning over multiple combinations, we find the best sensitivity for a product of the momenta of the 
second $p_T$-ordered tagging jet, $\vec{p}_{j_2}$, the Higgs reconstructed from the two $b$-jets 
$\vec{p}_{bb}$ and the photon $\vec{p}_{\gamma}$
\begin{align}
    \zeta = \frac{\vec{p}_\gamma \cdot (\vec{p}_{j_2} \times \vec{p}_{bb}) }{|\vec{p}_\gamma| |\vec{p}_{j_2}| |\vec{p}_{bb}| } \, .
            \label{eq:CPangle}
\end{align}
$CP$-odd operators create an asymmetry between the number of events with positive  and 
negative $\zeta$, which we denote by $N_{\zeta^+}$ and $N_{\zeta^-}$ respectively. 
In Fig.~\ref{fig:CPangle}, we compare the distributions of $\zeta$ in the SM 
with the ones in the EFT for rather extreme values of the Wilson coefficients. 
While $\zeta$ is symmetric for the SM case, the EFT clearly introduces 
an asymmetry in it.
\begin{figure}[t!]
\centering
\includegraphics[width=.48\textwidth]{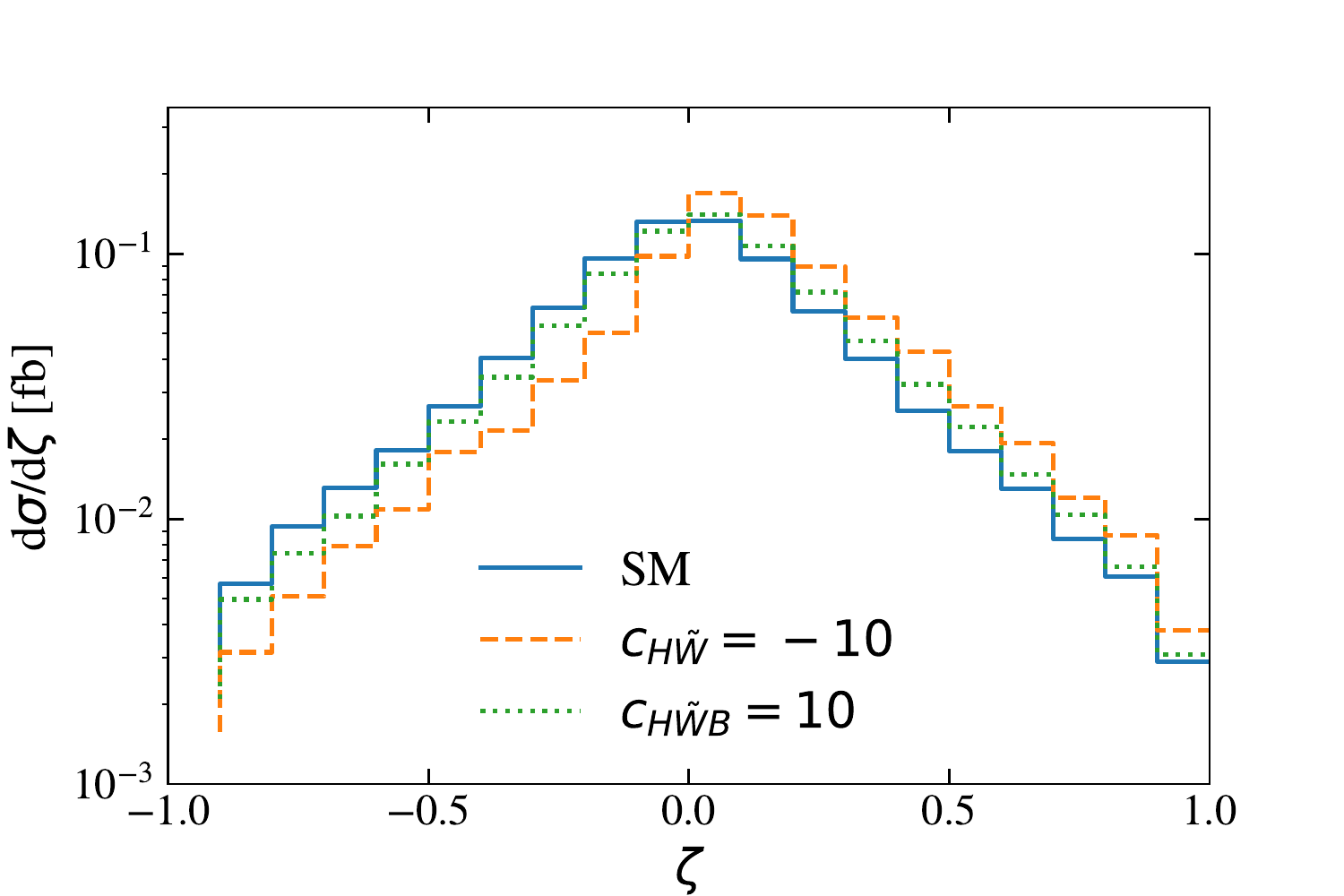}
\caption{Distribution of the $CP$ angle $\zeta$ in the SM and including the interference with the EFT.}
\label{fig:CPangle}
\end{figure}

As discussed before, there is no contribution to the total cross section from the interference of the $CP$-odd EFT with the $CP$-even SM. 
Therefore, rather than studying the event numbers $N_\zeta^\pm$ directly, 
we examine their normalized asymmetries
\begin{align}
    A_\zeta &= \frac{N_{\zeta}^+ - N_{\zeta}^- }{N_{\zeta}^+  + N_{\zeta}^- } , \quad
     \quad  A_\zeta^\text{SM} = 0 .
    \label{eq:asymmetry}
\end{align}

After the baseline cuts of Eq.~\eqref{eq:baseline_cuts}, we 
can parametrize the asymmetry in terms of the Wilson coefficients as 
\begin{align}
A_\zeta = 10^{-3}\cdot\left(\frac{1\tev}{\Lambda}\right)^2\cdot \Big[ -39 \, c_{H\tilde{W}} 
	+ 12 \, c_{H\tilde{W}B} \Big].
\label{eq:asym_para_baseline}
\end{align}
Taking into account only the statistical uncertainty, this allows us to constrain
the Wilson coefficients $c_{H\tilde{W}}$ and $c_{H\tilde{W}B}$ to 
\begin{align}
\frac{c_{H\tilde{W}}}{\Lambda^2} < \frac{1.1}{\tev^2}\qquad  
\frac{c_{H\tilde{W}B}}{\Lambda^2} < \frac{3.6}{\tev^2}\quad \text{ at 95\% CL}.
\label{eq:limits_baseline}
\end{align}
\begin{figure}[tb]
\centering
\includegraphics[width=.48\textwidth]{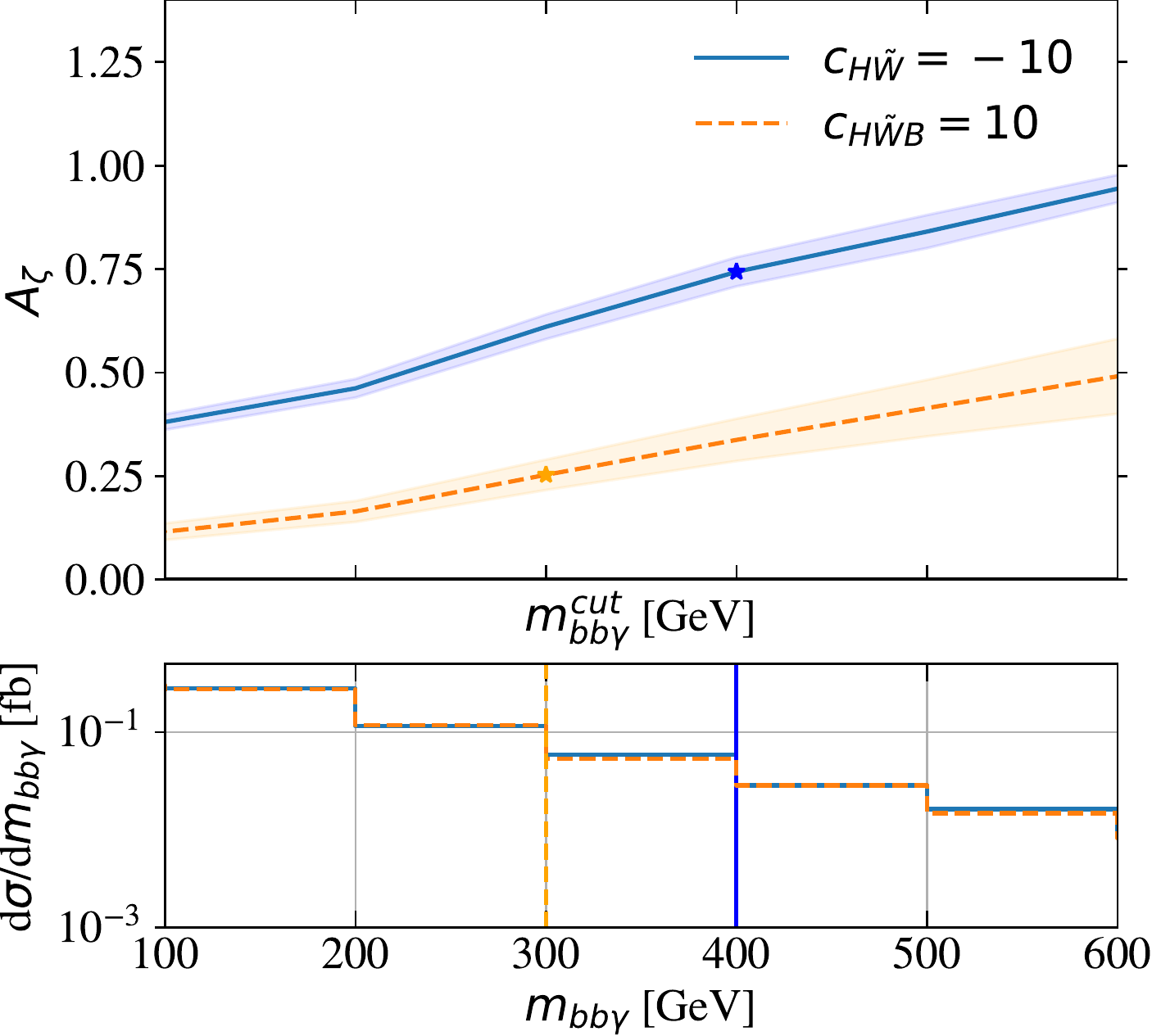}
\caption{Dependence of the asymmetry~$A_\zeta$ on a cut on the invariant mass of the 
Higgs-photon pair~$m_{bb\gamma} > m_{bb\gamma}^\text{cut}$. The shaded 
band represents the statistical uncertainty on the asymmetry assuming an integrated luminosity 
of~$3\,\iab$. The asterisk highlights the optimal cut on the invariant mass for 
the given Wilson coefficient. 
In the lower panel we show the distribution of the cross section as a 
function of the $m_{bb\gamma}$ invariant mass. }
\label{fig:mbba_vs_CPasym}
\end{figure}

In principle, the magnitude of the asymmetry~$A_\zeta$ depends on the kinematic region
selected by our cuts, 
because the relative contributions of different diagrams can be enhanced in different regions.
As an example, we display the dependence of the $CP$ 
asymmetry on a cut on the invariant mass of the Higgs-photon pair~$m_{bb\gamma}$ 
in Fig.~\ref{fig:mbba_vs_CPasym}. 
The asymmetry clearly rises with an increasing cut on $m_{bb\gamma}$.\footnote{In our basis and assuming Lorentz gauge for the gauge bosons, 
the direct effective $WWh\gamma$ coupling fills the tails of the $m_{bb\gamma}$ distribution more efficiently 
than the dimension-six $WWh$ interaction, i.e.~the $WWh\gamma$ coupling becomes more relevant at high 
$m_{bb\gamma}$.}
However, as the cross section drops quickly with $m_{bb\gamma}$ as displayed 
in the lower panel of Fig.~\ref{fig:mbba_vs_CPasym}, 
the statistical uncertainty depicted by the shaded band around the asymmetry curve
blows up rapidly. 
Therefore, the significance of the  asymmetry measurement
is a trade-off between selecting a signal region with a large asymmetry 
and keeping the measurement inclusive to reduce statistical uncertainties.

Assuming an optimal cut on the invariant mass of the 
Higgs-photon pair~$m_{bb\gamma}$ 
(inclusive for $c_{H\tilde{W}}$ and $m_{bb\gamma}>300 \,\gev$ for $c_{H\tilde{W}B}$)
we can improve the limits presented 
in Eq.~\eqref{eq:limits_baseline} to 
\begin{align}
\frac{c_{H\tilde{W}}}{\Lambda^2} < \frac{1.1}{\tev^2} \qquad 
\frac{c_{H\tilde{W}B}}{\Lambda^2} < \frac{3.1}{\tev^2} \quad \text{ at 95\% CL}.
\label{eq:limits_mbba}
\end{align}
We can compare our results with the limits from a global fit of the 
Higgs sector including WBF without an extra photon~\cite{Bernlochner:2018opw}, which for an integrated luminosity of $3\,\iab$ are quoted as $|\frac{c_{H\tilde{W}}}{\Lambda^2}|< \frac{1.2}{\tev^2}$ and 
$|\frac{c_{H\tilde{W}B}}{\Lambda^2}| < \frac{1.5}{\tev^2}$.\footnote{The quoted limits come from a global fit of the operators 
$\ope_{H\tilde{W}}$, $\ope_{H\tilde{W}B}$, 
$\ope_{H\tilde{G}}$ and $\ope_{H\tilde{B}}$. 
Since the limits on the Wilson coefficients 
of $\ope_{H\tilde{G}}$ and $\ope_{H\tilde{B}}$ stem mostly from gluon fusion Higgs production 
and the decay $h \rightarrow ZZ \rightarrow 4 \ell$ the limits 
on $c_{H\tilde{W}}$ and $c_{HB\tilde{W}}$
in a one-parameter fit 
should not significantly differ from the ones of a global fit. }
We would like to stress, though, our results for the limits are based on a comparison of SM $h\gamma$ production vs.\ the effect of SMEFT operators, and we did not include systematic uncertainties which we assume are larger for WBF+$\gamma$ than for WBF production alone.

Although our simplified analysis of WBF~$h\gamma$  does not clearly outperform the reference, the comparison underlines that a combination of 
our signal process with other signatures probing the same dimension-six operators 
is worth the effort, as it tests the underlying paradigms of the EFT construction and may lift degeneracies in a global fit. 

\section{Conclusions and Outlook}\label{sec:conclusions}

In this paper, we presented the prospects of measuring the $b$-Yukawa coupling or, conversely, the signal strength $\mu$ of the associated Higgs boson plus photon production in weak boson fusion with the Higgs boson decaying to bottom quarks at the LHC and the HL-LHC upgrade. 
The intricate kinematics of the five-particle final state render WBF~$h\gamma$ production 
a prime candidate for the application of multivariate analysis techniques. 
In fact, the resulting limit on the signal 
strength is narrowed from $\mu<1.1$ in a cut-and-count approach
to $\mu<0.8$ using a BDT analysis 
for the luminosity of the current ATLAS search $\lag_\text{int} =30.6\,\ifb$.
Tighter limits can be set with larger data sets, reaching 
$\mu<0.25$ and $\mu<0.1$ with $\lag_\text{int} =300\,\ifb$ 
and $\lag_\text{int} =3000\,\ifb$, respectively, neglecting 
systematic uncertainties.
This clearly indicates the possibility of observing this process at higher 
luminosity.

We also investigate the potential of this signature 
to limit non-Standard-Model-couplings, parametrized 
in the SMEFT framework.
Due to the presence of the additional photon compared 
to Higgs boson production in WBF only, the $CP$-even operators 
are four-boson operators and, thus, lack the addditional 
momentum dependence of the three-boson vertices. 
Hence, we do not expect competitive limits on them. 

The $CP$-odd operators, on the other hand, can be 
most meaningfully measured using asymmetries. 
Using $A_\zeta$ from Eq.\ \ref{eq:asym_para_baseline} we extract the following limits 
\begin{align}
  \frac{c_{H\tilde{W}}}{\Lambda^2} < \frac{1.1}{\tev^2} \qquad 
  \frac{c_{H\tilde{W}B}}{\Lambda^2} < \frac{3.1}{\tev^2}
\end{align}
at 95\% CL with the full HL-LHC dataset of $3\,\iab$. 
Again, as the measurement of this signature will 
be statistically limited we have ignored systematic 
uncertainties.

\section*{Acknowledgements}
We thank Filitsa Kougioumtzoglou for collaborating in the early stages of these studies.
Our work is supported by the UK Science and Technology Facilities Council (STFC) under grant  ST/P001246/1. 
FK and MS are acknowledging support from the European Union's Horizon 2020 research and innovation programme as part of the Marie Sklodowska-Curie Innovative Training Network MCnetITN3 (grant agreement no. 722104).
MS is funded by the Royal Society through a University Research Fellowship, and FK gratefully acknowledges support by the Wolfson Foundation and the Royal Society under award RSWF\textbackslash{}R1\textbackslash{}191029.

\bibliographystyle{elsarticle-num}
\bibliography{literature}

\end{fmffile}
\end{document}